\documentclass[a4paper]{article}

\usepackage{RR}
\RRNo{6366}

\usepackage{graphicx} 
\usepackage{algorithm}
\usepackage[isolatin]{inputenc}
\usepackage[OT1]{fontenc}                       
\usepackage{latexsym}
\usepackage{epsfig}
\usepackage{fancybox}

\newenvironment{proof}{{\bf Proof. } }{{\hfill $\Box$}\vspace{.5pc}}

\newtheorem{theorem}{Theorem}
\newtheorem{definition}{Definition}
\newtheorem{property}{Property}
\newtheorem{lemma}{Lemma}
\newtheorem{remark}{Remark}
\newtheorem{corollary}{Corollary}

\newtheorem{observation}{Observation}
\newtheorem{notation}{Notation}

\newcommand{\Prob}{\mathtt{Prob}}
\newcommand{\Det}{\mathtt{Det}}
\newcommand{\Rand}{\mathtt{Rand}}
\newcommand{\tokenh}{\mathtt{To\-ken\-Hol\-ders}}
\newcommand{\LCS}{\mathcal{LCSET}}
\newcommand{\voisin}{Neig}
\newcommand{\ROOT}{\mathtt{Root}}
\newcommand{\leader}{isLeader}
\newcommand{\LC}{\mathcal{LC}}
\newcommand{\ppath}{\mathtt{ParPath}}
\newcommand{\predpath}{\mathtt{PredPath}}
\newcommand{\mtd}{\mathtt{MTD}}
\newcommand{\C}{\mathcal{C}}
\newcommand{\A}{\mathtt{A}}
\newcommand{\Syst}{\mathcal{S}}
\newcommand{\I}{\mathcal{I}}
\newcommand{\Legi}{\mathcal{L}}
\newcommand{\Spec}{\mathcal{SP}}
\newcommand{\PROTO}{\mathcal{P}}
\newcommand{\Trans}{\mathtt{Trans}}

\RRNo{1}

\RRdate{November 2007}

\RRtitle{Stabilisation faible \emph{vs.} Auto-Stabilisation \emph{vs.} Stabilisation probabiliste}
\RRetitle{Weak \emph{vs.} Self \emph{vs.} Probabilistic Stabilization}

\titlehead{Weak \emph{vs.} Self \emph{vs.} Probabilistic}

\RRauthor{Stéphane Devismes\thanks{CNRS, Université Paris-Sud, France}
\and
Sébastien Tixeuil\thanks{Université Paris 6, LIP6-CNRS \& INRIA, France}
\and
Masafumi Yamashita\thanks{CSCE, Kyushu University, Japan}}

\RRresume{L'auto-stabilisation est une propriété forte qui assure qu'un réseau retrouve toujours un comportement correct quel que soit son état initial. Des propriétés plus faibles que l'auto-stabilisation ont été définies pour résoudre des résultats d'impossibilité: l'auto-stabilisation probabiliste garantit uniquement une convergence probabiliste vers un comportement correct; la stabilisation faible garantit simplement une possibilité de convergence à partir de n'importe quel état du système.

Dans cet article, nous nous intéressons aux puissances d'expression relatives de la stabilisation faible, de l'auto-stabilisation déterministe et de l'auto-stabilisation probabiliste. Nous prouvons qu'en pratique la stabilisation faible a réellement un pouvoir d'expression plus fort que l'auto-stabilisation déterministe ({\em i.e.}, elle permet de résoudre plus de problèmes que l'auto-stabilisation dé\-ter\-mi\-nis\-te). Ensuite, nous affinons des résultats antérieurs sur la stabilisation faible pour prouver que du point de vue pratique un protocole faiblement stabilisant déterministe peut être transformé en un protocole auto-stabilisant probabiliste. Ce résultat démontre l'intérêt pratique de la stabilisation faible puisque de tels algorithmes sont plus simples à écrire et à prouver que leurs équivalents auto-stabilisants (probabilistes).} 

\RRabstract{Self-stabilization is a strong property that guarantees that a network always resume correct behavior starting from an arbitrary initial state. Weaker guarantees have later been introduced to cope with impossibility results: probabilistic stabilization only gives probabilistic convergence to a correct behavior. Also, weak stabilization only gives the possibility of convergence.

In this paper, we investigate the relative power of weak, self, and probabilistic stabilization, with respect to the set of problems that can be solved. We formally prove that in that sense, weak stabilization is strictly stronger that self-stabilization. Also, we refine previous results on weak stabilization to prove that, for practical schedule instances, a deterministic weak-stabilizing protocol can be turned into a probabilistic self-stabilizing one. This latter result hints at more practical use of weak-stabilization, as such algorthms are easier to design and prove than their (probabilistic) self-stabilizing counterparts.}

\RRmotcle{Syst\`{e}mes distribu\'{e}s, Algorithme distribu\'{e}, Auto-stabilisation, Stabilisation faible, Auto-stabilisation probabiliste}
\RRkeyword{Distributed systems, Distributed algorithm, Self-stabilization, Weak-sta\-bi\-li\-za\-tion, Probabilistic self-stabilization}

\RRprojet{Grand large}
\RRtheme{\THNum}
\URFuturs

\begin{document}

\makeRR

\section{Introduction}

Self-stabilization~\cite{Dij74,citeulike:976108} is a versatile technique to withstand \emph{any} transient fault in a distributed system or network. Informally, a protocol is self-stabilizing if, starting from \emph{any} initial configuration, \emph{every} execution eventually reaches a point from which its behavior is correct. Thus, self-stabilization makes no hypotheses on the nature or extent of faults that could hit the system, and recovers from the effects of those faults in a unified manner.

Such versatility comes with a cost: self-stabilizing protocols can make use of a large amount of resources, may be difficult to design and to prove, or could be unable to solve some fundamental problems in distributed computing. To cope with those issues, several weakened forms of self-stabilization have been investigated in the literature. \emph{Probabilistic self-stabilization}~\cite{DBLP:conf/podc/IsraeliJ90} weakens the guarantee on the convergence property: starting from any initial configuration, an execution reaches a point from which its behavior is correct with probability $1$. \emph{Pseu\-do-sta\-bi\-li\-za\-tion}~\cite{BGM93} relaxes the notion of ``point'' in the execution from which the behavior is correct: every execution simply has a suffix that exhibits correct behavior, yet the time before reaching this suffix is unbounded. The notion of \emph{$k$-sta\-bi\-li\-za\-tion}~\cite{DBLP:conf/podc/BeauquierGK98} prohibits some of the configurations from being possible initial states, and assumes that an initial configuration may only be the result of $k$ faults (the number of faults being defined as the number of process memories to change to reach a correct configuration). Finally, the \emph{weak-sta\-bi\-li\-zat\-ion}~\cite{Gouda01} stipulates that starting from \emph{any} initial configuration, \emph{there exists} an execution that eventually reaches a point from which its behavior is correct. 

Probabilistic self-stabilization was previously used to reduce resource consumption~\cite{Her92} or to solve problems that are known to be impossible to solve in the classical deterministic setting~\cite{GT07}, such as graph coloring, or token passing. Also, it was shown that the well known alternating bit protocol is pseudo-stabilizing, but not self-stabilizing, establishing a strict inclusion between the two concepts. For the case of $k$-stabilization, \cite{GT02c,T07c} shows that if not all possible configurations are admissible as initial ones, several problems that can not be solved in the self-stabilizing setting (\emph{e.g.} token passing) can actually be solved in a $k$-stabilizing manner. As for weak-stabilization, it was only shown~\cite{Gouda01} that a sufficient condition on the scheduling hypotheses makes a weak-stabilizing solution self-stabilizing. 

From a problem-centric point of view, the probabilistic, pseudo, and $k$ variants of stabilization have been demonstrated strictly more powerfull that classical self-stabilization, in the sense that they can solve problems that are otherwise unsolvable. This comforts the intuition that they provide weaker guarantees with respect to fault recovery. In contrast, no such knowledge is available regarding weak-stabilization.  

In this paper, we address the latter open question, and investigate the power of weak-stabilization. Our contribution is twofold: \emph{(i)} we prove that from a problem centric point of view, weak-stabilization is stronger than self-stabilization (both for static problems, such as leader election, and for dynamic problems, such as token passing), and \emph{(ii)} we show that there exists a strong relationship between deterministic weak-stabilizing algorithms and probabilistic self-stabilizing ones. Practically, any deterministic weak-stabilizing protocol can be transformed into a probabilistic self-stabilizing protocol performing under a probabilistic scheduler, as we demonstrate in the sequel of the paper. This results has practical impact: it is much easier to design and prove a weak-stabilizing solution than a probabilistic one; so if new simple weak-stabilizing solutions appear in the future, our scheme can automatically make them self-stabilizing in the probabilistic sense.  

The remaining of the paper is organized as follows. In the next section we present the model we consider in this paper. In Section \ref{sect:3}, we propose weak-stabilizing algorithms for problems having no deterministic self-stabilizing solutions. In Section \ref{sect:2}, we show that under some scheduling assumptions, a weak-stabilizing system can be seen as a probabilistic self-stabilizing one.

\section{Model}\label{sect:model}

\noindent\textbf{Graph Definitions.} An \emph{undirected graph} $G$ is a couple $(V$,$E)$ where $V$ is a set of $N$ \emph{nodes} and $E$ is a set of \emph{edges}, each edge being a pair of distinct nodes. Two nodes $p$ and $q$ are said to be \emph{neighbors} \emph{iff} $\{p$,$q\} \in E$. $\Gamma_p$ denotes the set of $p$'s neighbors. $\Delta_p$ denotes the \emph{degree} of $p$, \emph{i.e.}, $|\Gamma_p|$. By extention, we denote by $\Delta$ the degree of $G$, \emph{i.e.}, $\Delta$ $=$ max($\{\Delta_p$, $p \in V\}$). 

A \emph{path} of lenght $k$ is a sequence of nodes $p_0$, \dots, $p_k$ such that $\forall i$, $0 \leq i < k$, $p_i$ and $p_{i+1}$ are neighbors. The path $\mathcal P = p_0$, \dots, $p_k$ is said \emph{elementary} if $\forall i$,$j$, $0 \leq i < j \leq k$, $p_i \neq p_j$. A path $\mathcal P = p_0$, \dots, $p_k$ is called \emph{cycle} if $p_0$, \dots, $p_{k-1}$ is elementary and $p_0 = p_k$. We call \emph{ring} any graph isomorph to a cycle.

An undirected graph $G = (V$,$E)$ is said \emph{connected} \emph{iff} there exists a path in $G$ between each pair of distinct nodes. The \emph{distance} between two nodes $p$ and $q$ in an undirected connected graph $G = (V$,$E)$ is the length of the smallest path between $p$ and $q$ in $G$. We denode the distance between $p$ and $q$ by $d(p$,$q)$. The diameter $D$ of $G$ is equal to max($\{d(p$,$q)$, $p \in V \wedge q \in V\}$). The eccentricity of a node $p$, noted $ec(p)$, is equal to max($\{d(p$,$q)$, $q \in V\}$). A node $p$ is a \emph{center} of $G$ if $\forall q \in V$, $ec(p) \leq ec(q)$.

We call \emph{tree} any undirected connected acyclic graph. In a tree graph, we distinghish two types of nodes: the \emph{leaves} (\emph{i.e.}, any node $p$ such that $\Gamma_p = 1$) and the \emph{internal nodes} (\emph{i.e.}, any node $p$ such that $\Gamma_p > 1$). Below, we recall a well-known result about the centers in the trees.

\begin{property}[\cite{BH90}]\label{prop:center}
A tree has a unique center or two neighboring centers. 
\end{property}

\noindent\textbf{Distributed Systems. } A \emph{distributed system} is a finite set of communicating state machines called \emph{processes}. We represent the \emph{communication network} of a distributed system by the undirected connected graph $G = (V$,$E)$ where $V$ is the set of $N$ processes and $E$ is a set of edges such that $\forall p$,$q \in V$, $\{p$,$q\} \in E$ \emph{iff} $p$ and $q$ can directly communicate together. Here, we consider \emph{anonymous} distributed systems, \emph{i.e.}, the processes can only differ by their degrees.  We assume that each process can distinguish all its neighbors using \emph{local indexes}, these indexes are stored in $Neig_p$. For sake of simplicity, we assume that $\voisin_p = \{0$, \dots, $\Delta_p -1\}$. In the following, we will indifferently use the \emph{label} $q$ to designate the process $q$ or the local index of $q$ in the code of some process $p$. 

The communication among neighboring processes is carried out using a \emph{finite} number of \emph{shared variables}. Each process holds its own set of shared variables where it is the only able to write but where each of its neighbors can read. The \emph{state} of a process is defined by the values of its variables. A \emph{configuration} of the system is an instance of the state of its processes. A process can change its state by executing its \emph{local algorithm}. The local algorithm executed by each process is described by a finite set of guarded actions of the form: $\langle label \rangle :: \langle guard \rangle \to \langle statement \rangle$. The guard of an action at Process $p$ is a boolean expression involving some variables of $p$ and its neighbors. The statement of an action of $p$ updates some variables of $p$. An action can be executed only if its guard is satisfied. We assume that the execution of any action is \emph{atomic}. An action of some process $p$ is said enabled in the configuration $\gamma$ \emph{iff} its guard is $true$. By extention, $p$ is said \emph{enabled} in $\gamma$ \emph{iff} at least one of its action is enabled in $\gamma$. 

We model a distributed system as a \emph{transition system} $\Syst = (\C$,$\mapsto$,$\I)$ where $\C$ is the set of system configuration, $\mapsto$ is a binary transition relation on $\C$, and $\I \subseteq \C$ is the set of initial configurations. An \emph{execution} of $\Syst$ is a \emph{maximal} sequence of configurations $\gamma_0$, \dots, $\gamma_{i-1}$, $\gamma_i$, \dots\ such that $\gamma_0 \in \I$ and $\forall i > 0$, $\gamma_{i-1} \mapsto \gamma_i$ (in this case, $\gamma_{i-1} \mapsto \gamma_i$ is referred to as a \emph{step}). Any configuration $\gamma$ is said \emph{terminal} if there is no configuration $\gamma'$ such that $\gamma \mapsto \gamma'$. We denote by $\gamma \leadsto \gamma'$ the fact that $\gamma'$ is \emph{reachable} from $\gamma$, \emph{i.e.}, there exists an execution starting from $\gamma$ and containing $\gamma'$. 

A \emph{scheduler} is a predicate over the executions. In any execution, each step $\gamma \mapsto \gamma'$ is obtained by the fact that a \emph{non-empty} subset of enabled processes \emph{atomically execute} an action. This subset is chosen according to the scheduler. A scheduler is said \emph{central} \cite{Dij74} if it chooses \emph{one} enabled process to execute an action in any execution step. A scheduler is said \emph{distributed} \cite{BGM89} if it chooses \emph{at least one} enabled process to execute an action in any execution step. A scheduler may also have some \emph{fairness} properties (\cite{citeulike:976108}). A scheduler is \emph{strongly fair} (the strongest fairness assumption) if every process that is enabled \emph{infinitely often} is eventually chosen to execute an action. A scheduler is \emph{weakly fair} if every \emph{continuously} enabled process is eventually chosen to execute an action. Finally, the \emph{proper} scheduler is the weakest fairness assumption: it can forever prevent a process to execute an action except if it is the only enabled process. As the strongly fair scheduler is the strongest fairness assumption, any problem that cannot be solved under this assumption cannot be solved for all fairness assumptions. In contrast, any algorithm working under the proper scheduler also works for all fairness assumptions.

We call $\mathtt{P}\mbox{-}\mathtt{variable}$ any variable $v$ such that there exists a statement of an action where $v$ is randomly assigned. Any variable that is not a $\mathtt{P}\mbox{-}\mathtt{variable}$ is called $\mathtt{D}\mbox{-}\mathtt{variable}$. Each random assignation of the $\mathtt{P}\mbox{-}\mathtt{variable}$ $v$ is assumed to be performed using a random function $\Rand_v$ which returns a value in the domain of $v$. A system is said \emph{probabilistic} if it contains at least one $\mathtt{P}\mbox{-}\mathtt{variable}$, otherwise it is said \emph{deterministic}. Let $\Syst = (\C$,$\mapsto$,$\I)$ be a probabilistic system. Let $\mathtt{Enabled}(\gamma)$ be the set of processes that are enabled in $\gamma \in \C$. $\Syst$ satisfies: for any subset $\mathtt{Sub}(\gamma) \subseteq \mathtt{Enabled}(\gamma)$, the sum of the probabilities of the execution steps determined by $\gamma$ and $\mathtt{Sub}$ is equal to 1.

\bigskip

\noindent\textbf{Stabilizing Systems.} Let $\Syst = (\C$,$\mapsto$,$\I)$ be a system such that $\C = \I$ (\emph{n.b.}, in the following any system $\Syst = (\C$,$\mapsto$,$\I)$ such that $\C = \I$ will be simply denoted by $\Syst = (\C$,$\mapsto)$). Let $\Spec$ be a specification, \emph{i.e.}, a particular predicate defined over the executions of $\Syst$.

\begin{definition}[Deterministic Self-Stabilization \cite{Dij74}]\label{def:self} $\Syst$ is \emph{de\-ter\-mi\-nisti\-cal\-ly} \emph{self-sta\-bi\-li\-zing} for $\Spec$ if there exists a non-empty subset of $\C$, noted $\Legi$, such that: \emph{(i)} Any execution of $\Syst$ starting from a configuration of $\Legi$ always satisfies $\Spec$ (\emph{Strong Closure Property}), and \emph{(ii)} Starting from any configuration, \emph{any execution} of $\Syst$ reaches in a finite time a configuration of $\Legi$ (\emph{Certain Convergence Property}).
\end{definition}

\begin{definition}[Pro\-ba\-bi\-li\-stic Self-Sta\-bi\-li\-za\-tion \cite{DBLP:conf/podc/IsraeliJ90}]\label{def:selfprob}
$\Syst$ is \emph{pro\-ba\-bi\-li\-sti\-cal\-ly} \emph{self-sta\-bi\-li\-zing} for $\Spec$ if there exists a non-empty subset of $\C$, noted $\Legi$, such that: \emph{(i)} Any execution of $\Syst$ starting from a configuration of $\Legi$ always satisfies $\Spec$ (\emph{Strong Closure Property}), and \emph{(ii)} Starting from any configuration, any execution of $\Syst$ reaches a configuration of $\Legi$ \emph{with Probability 1} (\emph{Probabilistic Convergence Property}).
\end{definition}

\begin{definition}[Deterministic Weak-Stabilization \cite{Gouda01}]\label{def:weak}
$\Syst$ is \emph{deterministically weak-sta\-bi\-li\-zing} for $\Spec$ if there exists a non-empty subset of $\C$, noted $\Legi$, such that: \emph{(i)} Any execution of $\Syst$ starting from a configuration of $\Legi$ always satisfies $\Spec$ (\emph{Strong Closure Property}), and \emph{(ii)} Starting from any configuration, \emph{there always exists an execution} that reaches a configuration of $\Legi$ (\emph{Possible Convergence Property}).
\end{definition}

\noindent Note that the configurations from which $\Syst$ always satisfies $\Spec$ ($\Legi$) are called \emph{legitimate configurations}. Conversely, every configuration that is not legitimate is \emph{illegitimate}.

\section{From Self to Weak Stabilization}\label{sect:3}

In this section, we exhibit two problems that can not be solved by a deterministic self-stabilizing protocol, yet admit surprisingly simple deterministic weak-stabilizing ones. Thus, from a problem-centric point of view, weak-stabilization is stronger than self-stabilization.
This result is mainly due to the fact that a given scheduler is appreciated differently when we consider self or weak stabilization. In the self-stabilizing setting, the scheduler is seen as an \emph{adversary}: the algorithm must work properly despite the "bad behavior" of the scheduler. Indeed, it is sufficient to exhibit an execution that satisfies the scheduler predicate yet prevents the algorithm from converging to a legitimate configuration to prove the absence of self-stabilization. 
Conversely, in weak-stabilization, the scheduler can be viewed as a \emph{friend}: to prove the property of weak-stabilization, it is sufficient to show that, for any configuration $\gamma$, there exists an execution starting from $\gamma$ that satisfies the scheduler predicate and converges. As a matter of fact, the effect of the scheduler is reversed in weak and self stabilization: the strongest the scheduler is (\emph{i.e.} the more executions are included in the scheduler predicate), the easier the weak-stabilization can be established, but the harder self-stabilization is. 

When the scheduler is \emph{synchronous}~\cite{DBLP:journals/ipl/Herman90} (\emph{i.e.}, a scheduler that chooses \emph{every} enabled process at each execution step) the notions of deterministic weak-stabilization and deterministic self-stabilization are equivalent, as proved in the following.

\begin{theorem}Under a syn\-chro\-nous sche\-du\-ler, an algorithm is deterministically weak-sta\-bi\-li\-zing \emph{iff} it is al\-so de\-ter\-mi\-ni\-sti\-cal\-ly self-sta\-bi\-li\-zing.
\end{theorem}
\begin{proof}~

{\bf If.} Consider algorithm $\PROTO$ that is deterministically weak-stabilizing under a synchronous scheduler. First, $\PROTO$ satisfies the \emph{strong closure} property. It remains then to show that $\PROTO$ satisfies the \emph{certain convergence} property. 

By Definition \ref{def:weak}, starting from any configuration $\gamma$, there exists an execution of $\PROTO$ that converges to a legitimate configuration. Now, under a synchronous scheduler, there is an unique execution starting from $\gamma$ because $\PROTO$ is deterministic. Hence, $\PROTO$ trivially satisfies the following assertion "starting from any configuration, any execution of $\PROTO$ converges to a legitimate configuration under a synchronous scheduler" (the \emph{certain convergence} property).

{\bf Only If.} By Definition, any deterministic self-stabilizing algorithm is also a deterministic weak-stabilizing algorithm under the same scheduler.
\end{proof}

We now exhibit two examples of problems that admit weak-stabilizing solutions but no self-stabilizing ones: the token passing and the leader election.

\subsection{Token Circulation}
\label{sub:TC}

In this subsection, we consider the problem of Token Circulation in a unidirectional ring, with a strongly fair distributed scheduler. This problem is one of the most studied problems in self-stabilization, and is often regarded as a ``benchmark'' for new algorithms and concepts. The consistent direction is given by a constant local pointer $Pred$: for any process $p$, $Pred_p$ designates a neighbor $q$ as the \emph{predecessor} (resp. $p$ is the \emph{successor} of $q$) in such way that $q$ is the predecessor of $p$ \emph{iff} $p$ is not the predecessor of $q$. 

\begin{definition}[Token Circulation]
The \emph{token circulation} problem consists in circulating a single token in the network in such way that every process holds the token infinitely often. 
\end{definition}

In \cite{DBLP:journals/ipl/Herman90}, Herman shows, using a previous result of Angluin \cite{A80}, that the deterministic self-stabilizing token circulation is impossible in anonymous networks because there is no ability to break symmetry. We now show that, contrary to deterministic self-stabilization, deterministic weak-stabilizing token circulation under distributed strongly fair scheduler exists in an anonymous unidirectional ring.

Our starting point is the $(N-1)$-fair algorithm of Beauquier \emph{et al.} proposed in \cite{DBLP:journals/dc/BeauquierGJ07} (presented as Algorithm \ref{algo:2}). We show that Algorithm~\ref{algo:2} is actually a deterministic weak-stabilizing token circulation protocol. Roughly speaking, $(N-1)$-fairness implies that in any execution, \emph{(i)} every process $p$ performs actions infinitely often, and \emph{(ii)} between any two actions of $p$, any other process executes at most $N-1$ actions. The memory requirement of Algorithm \ref{algo:2} is $\log (m_{N})$ bits per process where $m_{N}$ is the smallest integer not dividing $N$ (the ring size). Note that it is also shown in \cite{DBLP:journals/dc/BeauquierGJ07} that this memory requirement is minimal to obtain any probabilistic self-stabilizing token circulation under a distributed scheduler (such a probabilistic self-stabilizing token circulation can be found in \cite{DGT04j}). 

\begin{algorithm}\caption{Code for every process $p$\label{algo:2}}
\scriptsize
{\bf Variable:} $dt_p \in [0\dots m_{N} - 1]$\\
{\bf Macro:}\\
\begin{tabular}{lll}
$PassToken_p$ & $=$ & $dt_p \gets (dt_{Pred_p} + 1) \bmod m_{N}$
\end{tabular}\\
{\bf Predicate:}\\
\begin{tabular}{lll}
$Token(p)$ & $\equiv$ & $[dt_p \neq ((dt_{Pred_p} + 1) \bmod m_{N})]$
\end{tabular}\\
{\bf Action:}\\
\begin{tabular}{lllll}
$\mathtt{A}$ & $::$ & $Token(p)$ & $\to$ & $PassToken_p$
\end{tabular}
\end{algorithm}

A process $p$ maintains a single counter variable: $dt_p$ such that $dt_p \in [0\dots m_{N} - 1]$. This variable allows $p$ to know if it holds the token or not. Actually, a process $p$ holds a token \emph{iff} $dt_p \neq ((dt_{Pred_p} + 1) \bmod m_{N})$, \emph{i.e.}, \emph{iff} $p$ satisfies $Token(p)$. In this case, Action $\mathtt{A}$ is enabled at $p$. This action allows $p$ to pass the token to its \emph{successor}. 

\begin{figure*}[t]
\begin{center}
\epsfig{file=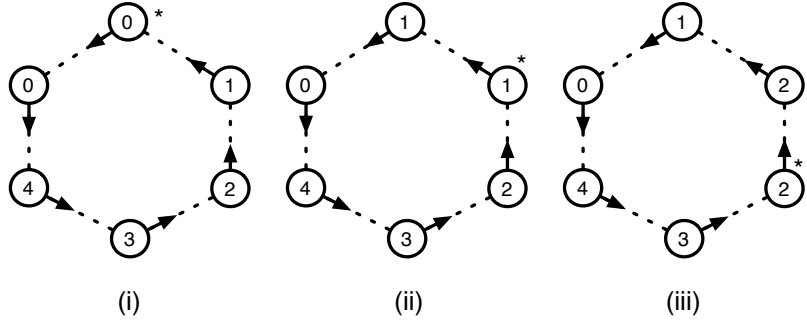, width=300pt, height=120pt}
\caption{Example of an execution starting from a legitimate configuration.\label{ex3}}
\end{center}
\end{figure*}

Figure \ref{ex3} depicts an execution of Algorithm \ref{algo:2} starting from a legitimate configuration, \emph{i.e.}, a configuration where there is exactly one process that satisfies Predicate $Token$. In the figure, the outgoing arrows represent the $Pred$ pointers and the integers represent the $dt$ values. In this example, the ring size $N$ is equal to 6. So, $m_{N} = 4$. In each configuration, the only process with an asterisk is the only token holder: by executing Action $\mathtt{A}$, it passes the token to its successor.

\begin{theorem}\label{theo:algo2w}
Algorithm \ref{algo:2} is a \emph{deterministic weak-stabilizing token passing} algorithm under a distributed strongly fair scheduler.
\end{theorem}
\begin{proof}
Given in the appendix (Section \ref{app:A}, page \pageref{app:A}).
\end{proof}

\subsection{Leader Election}
\label{sub:LE}

In this subsection, we consider anonymous tree-shaped networks and a distributed strongly fair scheduler. 

\begin{definition}[Leader Election]
The \emph{leader election} problem consists in distinguishing a unique process in the network. 
\end{definition}

We first prove that the leader election problem is impossible to solve in our setting in a self-stabilizing way.

\begin{theorem}
Assuming a distributed strongly fair scheduler, there is no deterministic self-stabilizing leader election algorithm in anonymous trees.
\end{theorem}
\begin{proof}
Consider a chain of four processes $P_1$, $P_2$, $P_3$, $P_4$ (a particular case of tree) and a synchronous execution (a possible behavior of a distributed strongly fair scheduler). Let us denote by $\langle S_1$,$S_2$,$S_3$,$S_4\rangle$ any configuration of the system we consider where $S_i$ ($i \in [1\dots 4]$) represents the local state of $P_i$. Let $\mathcal X$ be the subset of configurations such that $S_1 = S_4$ and $S_2 = S_3$ (note that $S_1 = S_2 = S_3 = S_4$ is a particular case of such configurations). Of course, in any configuration of $\mathcal X$, we cannot distinghish any leader. We now show that $\mathcal X$ is closed in a synchronous execution, which proves the impossibility of the deterministic self-stabilizing leader election.

Consider a configuration $\gamma = \langle a$,$b$,$b$,$a\rangle$ of the set $\mathcal X$. As we cannot distinghish any leader in $\gamma$, $\gamma$ must not be terminal. So, consider an arbitrary execution starting from $\gamma$ and let $\gamma'$ be the configuration that follows $\gamma$ in the execution. The three following cases are possible for the step $\gamma \mapsto \gamma'$:
\begin{itemize}
\item[-] Only $P_1$ and $P_4$ are enabled in $\gamma$. As the system is deterministic and the execution is synchronous, there only one possible step: $P_1$ and $P_4$ changes their local state in the same deterministical way. So, $S_1$ is still identical to $S_4$ in $\gamma'$, \emph{i.e.}, $\gamma' = \langle a$,$b'$,$b'$,$a\rangle$. 
\item[-] Only $P_2$ and $P_3$ are enabled in $\gamma$. As the system is deterministic and the execution is synchronous, there only one possible step: $P_2$ and $P_3$ changes their local state in the same deterministical way. So, $S_2$ is still identical to $S_3$ in $\gamma'$, \emph{i.e.}, $\gamma' = \langle a$,$b'$,$b'$,$a\rangle$.
\item[-] All processes are enabled. In this case, we trivially have $\gamma' = \langle a'$,$b'$,$b'$,$a'\rangle$.
\end{itemize}
Hence, $\gamma' \in \mathcal X$, which proves that $\mathcal X$ is closed.
\end{proof}

\begin{figure*}[t]
\begin{center}
\epsfig{file=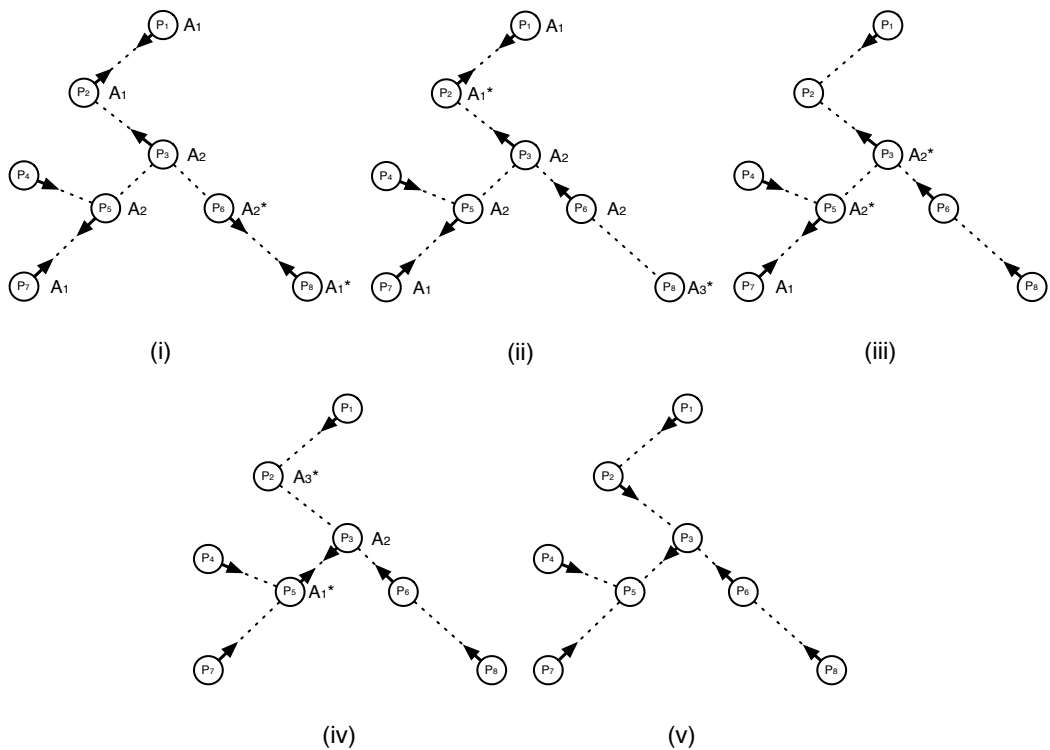, width=390pt, height=280pt}
\caption{Example of possible convergence.\label{ex2}}
\end{center}
\end{figure*}

We now provide two weak-stabilizing solutions for the same problem in the same setting, with different space complexities. Both solutions are more intuitive and simpler to design than self-stabilizing ones in slightly different settings. 

\bigskip

\noindent\textbf{A solution using $\log N$ bits.} A straighforward solution is to use the algorithm provided in \cite{DBLP:journals/siamcomp/BruellGKP99}. This algorithm uses $\log N$ bits and finds the centers of a tree network: starting from any configuration, the system reaches in a finite time a terminal configuration where any process $p$ satisfies a particular local predicate $Center(p)$ \emph{iff}\ $p$ is a center of the tree. From Property \ref{prop:center}, two cases are then possible in a terminal configuration: either a unique process satisfies $Center$ or two neighboring processes satisfy $Center$.

If there is only one process $p$ satisfying $Center(p)$, it is considered as the leader.

Now, assume that there are two neighboring processes $p$ and $q$ that satisfy $Center$. In this case, $p$ (resp. $q$) is able to locally detect that $q$ (resp. $p$) is the other center (see \cite{DBLP:journals/siamcomp/BruellGKP99} for details). So, we use an additional boolean $B$ to break the tie. If $B_p \neq B_q$, then the only center satisfying $B = true$ is considered as the leader. Otherwise, both $p$ and $q$ are enabled to execute $B \gets \neg B$. So, from any configuration where the two centers have been found but no leader is distinguished, this is always possible to reach a terminal configuration where a leader is distinghished in one step: if only one of the two centers moves.

\bigskip

\noindent\textbf{Another solution using $\log \Delta$ bits.} In this solution (Algorithm \ref{algo:1}), each process $p$ maintains a single variable: $Par_p$ such that $Par_p \in \voisin_p \cup \{\perp\}$. $p$ considers itself as the leader \emph{iff} $Par_p = \perp$. If $Par_p \neq \perp$, the \emph{parent} of $p$ is the neighbor pointed out by $Par_p$, conversely $p$ is said to be a \emph{child} of this process.

\begin{algorithm*}\caption{Code for any process $p$\label{algo:1}}
\scriptsize
{\bf Variable:} $Par_p \in \voisin_p \cup \{\perp\}$\\
{\bf Macro:}\\
\begin{tabular}{lll}
$Children_p$ & $=$ & $\{q \in \voisin_p$, $Par_q = p\}$
\end{tabular}\\
{\bf Predicates:}\\
\begin{tabular}{lll}
$\leader(p)$ & $\equiv$ & $(Par_p = \perp)$
\end{tabular}\\
{\bf Actions:}\\
\begin{tabular}{lllll}
 $\mathtt{A_1}$ & $::$ & $(Par_p \neq \perp) \wedge (|Children_p|=|\voisin_p|)$ & $\to$ & $Par_p \gets \perp$\\
 $\mathtt{A_2}$ & $::$ & $(Par_p \neq \perp) \wedge [\voisin_p \setminus (Children_p \cup \{Par_p\}) \neq \emptyset]$ & $\to$ & $Par_p \gets (Par_p + 1) \bmod \Delta_p$\\
$\mathtt{A_3}$ & $::$ & $(Par_p = \perp) \wedge (|Children_p|<|\voisin_p|)$ & $\to$ & $Par_p$ $\gets$ min$_{\prec_p} (\voisin_p \setminus Children_p)$
\end{tabular}
\end{algorithm*}

Algorithm \ref{algo:1} tries to reach a terminal configuration where: \emph{(i)} exactly one process $l$ is designated as the leader, and \emph{(ii)} all other processes $q$ point out using $Par_q$ their neighbor that is the closest from $l$. In other words, Algorithm \ref{algo:1} computes an arbitrary orientation of the network in a deterministic weak-stabilizing manner. 

Algorithm \ref{algo:1} uses the following strategy:
\begin{itemize}
\item[1.] If a process $p$ such that $Par_p \neq \perp$ is pointed out by all its neighbors, then this means that all its neighbors consider it as the leader. As a consequence, $p$ sets $Par_p$ to $\perp$ (Action $\mathtt{A_1}$), \emph{i.e.}, it starts to consider itself as the leader.
\item[2.] If a process $p$ such that $Par_p \neq \perp$ has a neighbor which is neither its parent nor one of its children, then this means that not all processes among $p$ and its neighbors consider the same process as the leader. In this case, $p$ changes its parent by simply incrementing its parent pointer modulus $\Delta_p$  (Action $\mathtt{A_2}$). Hence, from any configuration, it is always possible that all processes satisfying $Par \neq \perp$ eventually agree on the same leader.
\item[3.] Finally, if a process $p$ satisfies $Par_p = \perp$ and at least one of neighbor $q$ does not satisfy $Par_q = p$, then this means that $q$ considers another process as the leader. As a consequence, $p$ stops to consider itself as the leader by pointing out one of its non-child neighbor (Action $\mathtt{A_3}$).
\end{itemize} 

Figure \ref{ex2} depicts an example of execution of Algorithm \ref{algo:1} that converges. In the figure, the circles represent the processes and the dashed lines correspond to the neighboring relations. The labels of processes are just used for the ease of explanation. Then, if there is an arrow outgoing from process $P_i$, this arrow designates the neighbor pointed out by $Par_{P_i}$. In contrast, $Par_{P_i} = \perp$ holds if there is no arrow outgoing from process $P_i$. Any label $A_j$ beside a process $P_i$ means that Action $\mathtt{A_j}$ is enabled at $P_i$. Finally, some labels $A_j$ are sometime asterisked meaning that their corresponding actions is executed in the next step.

In initial configuration $(i)$, no process satisfies $Par = \perp$, \emph{i.e.}, no process consider itself as the leader. However, $P_1$, $P_2$, $P_7$, and $P_8$ are pointed out by all their respective neighbors. So, these processes are candidates to become the leader (Action $\mathtt{A_1}$). Also, note that $P_3$, $P_5$, and $P_6$ are enabled to execute Action $\mathtt{A_2}$: they have a neighbor that is neither their parent or one of their children. Finally, note that $P_4$ is in a stable local state. In the first step $(i) \mapsto (ii)$, $P_6$ and $P_8$ execute their enabled action: in $(ii)$, there is a unique leader ($P_8$) but it has no child, \emph{i.e.}, no other process agrees on its leadership. So $P_8$ is enabled to lose its leadership (Action $\mathtt{A_3}$). In $(ii) \mapsto (iii)$, $P_8$ looses its leadership (Action $\mathtt{A_3}$) but $P_2$ becomes a leader (Action $\mathtt{A_1}$). So, there is still a unique leader ($P_2$) in the configuration $(iii)$. In the step $(iii) \mapsto (iv)$, $P_3$ and $P_5$ change their parent to $P_5$ and $P_3$, respectively. As a consequence, Action $\mathtt{A_1}$ becomes enabled at $P_5$ in $(iv)$. However, $P_2$ is also enabled in $(iv)$ to lose its leadership (Action $\mathtt{A_3}$). In $(iv) \mapsto (v)$, $P_2$ and $P_5$ execute their respective enabled action and the system reach the terminal configuration $(v)$.

\begin{figure}
\begin{center}
\epsfig{file=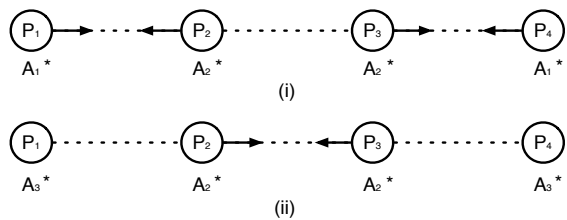, width=210pt, height=80pt}
\caption{Example of an execution that does not converge.\label{ex1}}
\end{center}
\end{figure}

Figure \ref{ex1} illustrates the fact that Algorithm \ref{algo:1} is deterministically weak-stabilizing but not deterministically self-stabilizing under a distributed scheduler (for all fairness assumptions). Actually Figure \ref{ex1} show that there is some infinite executions of Algorithm \ref{algo:1} that never converge. This example is quite simple: starting from the configuration $(i)$, if the execution is synchronous, the system reaches configuration $(ii)$ in one step, then we retreive configuration $(i)$ after two steps, and so on. This sequence can be repeated indefinitely. So, there is a possible execution starting from $(i)$ that never converges.

\begin{theorem}\label{theo:algo1w}
Algorithm \ref{algo:1} is a \emph{deterministic weak-stabilizing leader election} algorithm under a distributed strongly fair scheduler.
\end{theorem}
\begin{proof}
Given in the appendix (Section \ref{app:B}, page \pageref{app:B}).
\end{proof}

\section{From Weak to Probabilistic Stabilization}\label{sect:2}



In \cite{Gouda01}, Gouda shows that deterministic weak-stabilization is a ``good approximation'' of deterministic self-stabilization\footnote{This result has been proven for the central scheduler but it is easy to see that the proof also holds for any scheduler.} by proving the following theorem:

\begin{theorem}[\cite{Gouda01}]\label{theo:goudasf} Any deterministic weak-stabilizing system is also a deterministic self-stabilizing system if:
\begin{itemize}
\item[-] The system has a finite number of configurations, and
\item[-] Every execution satisfies the \emph{Gouda's strong fairness} assumption where \emph{Gouda's strong fairness} means that, for every transition $\gamma \mapsto \gamma'$, if $\gamma$ occurs infinitely often in an execution $e$, then $\gamma \mapsto \gamma'$ also appears infinitely often in $e$. 
\end{itemize}
\end{theorem}

From Theorem \ref{theo:goudasf}, one may conclude that deterministic weak-stabilization and deterministic self-stabilization are equivalent under the distributed strongly fair scheduler. This would contradict the results presented in Section~\ref{sect:3}. Actually, this is not the case: we prove in Theorem \ref{lem:gouda0-1} that the Gouda's strong fairness assumption is (strictly) stronger than the classical notion of strong fairness. 
A less ambiguous and more practical characterization of deterministic weak-stabilization is the following: under Gouda's strong fairness assumption, the scheduler does not behave as an adversary but rather as a probabilistic one (\emph{i.e.}, a deterministic weak-stabilizing system may never converge but if it is lucky, it converges). Hence, under a distributed randomized scheduler~\cite{DBLP:conf/sss/DasguptaGX07}, which chooses among enabled processes with a (possibly) uniform probability which are activated, any weak-stabilizing system converges with probability 1 despite an arbitrary initial configuration (Theorem \ref{gouda0-2}).

\begin{theorem}\label{lem:gouda0-1} The \emph{Gouda's strong fairness} is stronger than the \emph{strong fairness}.
\end{theorem}
\begin{proof}
As Algorithm \ref{algo:2} (page \pageref{algo:2}) is a deterministic weak-stabilizing token circulation with a finite number of configurations, it is also a deterministic self-stabilizing token circulation under the \emph{Gouda's strongly fairness assumption} (Theorem \ref{theo:goudasf}). We now show the lemma by exhibiting an execution of Algorithm \ref{algo:2} that does not converge under the central strongly fair scheduler (a similar counter-example can be also derived for a synchronous scheduler). 

Consider a ring of six processes $p_0$, \dots, $p_5$. Consider a configuration $\gamma_0$ where only $p_0$ and $p_3$ hold a token. Both $p_0$ and $p_3$ are enabled in $\gamma_0$. Assume that only $p_0$ passes its token in the step $\gamma_0 \mapsto \gamma_1$. In $\gamma_1$, $p_1$ and $p_3$ hold a token. Assume now that only $p_3$ passes its token in the step $\gamma_1 \mapsto \gamma_3$ and so on. It is straightforward that if the two tokens alternatively move at each step, then the execution never converges despite it respects the central strongly fair scheduler.
\end{proof}

We now show that the \emph{randomized scheduler} defined below is a notion that is, in some sense, equivalent to the \emph{Gouda's strong fairness}.

\begin{definition}[Randomized Scheduler \cite{DBLP:conf/sss/DasguptaGX07}]\label{def:dp}
A scheduler is said \emph{randomized} if it randomly chooses with a uniform probability the enabled processes that execute an action in each step. 
\end{definition}

Note that under a \emph{cen\-tral randomized scheduler}, in every step the unique process that executes an action is chosen with a uniform probability among the enabled processes. Similarly, under a \emph{distributed randomized scheduler}, in every step the processes (at least one) that executes an action are chosen with a uniform probability among the enabled processes.  

\begin{theorem}\label{gouda0-2}
Let $\PROTO$ be a deterministic algorithm having a finite number of configurations. $\PROTO$ is deterministically self-stabilizing under the Gouda's fairness assumption \emph{iff} $\PROTO$ is probabilistically self-stabilizing under a randomized scheduler.
\end{theorem}
\begin{proof}
 Let $\PROTO$ be a deterministic algorithm having a finite number of configurations.

{\bf If.} Assume that $\PROTO$ is deterministically self-stabilizing under the Gouda's fairness assumption. First, $\PROTO$ satisfies the \emph{strong closure} property. Hence, it remains to show that $\PROTO$ also satisfies the \emph{probabilistic convergence} property. 

Assume, by the contradiction, that there exists an execution $e$ of $\PROTO$ that do not converge with a probability 1 under a distributed randomized scheduler. As the number of configurations of $\PROTO$ is finite, there exists at least one configuration $\gamma_0$ that occurs infinitely often in $e$. Then, as $\PROTO$ is deterministically self-stabilizing under the \emph{Gouda's fairness assumption}, there exists an execution $\gamma_0$, $\gamma_1$, \dots, $\gamma_k$ such that $\gamma_k$ is a legitimate configuration. Now, as the scheduler is randomized, there is a strictly positive probability that $\gamma_0 \mapsto \gamma_1$ occurs starting from $\gamma_0$. Hence, $\gamma_0 \mapsto \gamma_1$ occurs with a probability 1 after a finite number of occurences of $\gamma_0$ in $e$ and, as a consequence, $\gamma_1$ occurs infinitely often (with the probability 1) in $e$. Inductively, it is then straightforward that $\forall i \in [1\dots k]$, $\gamma_i$ occurs infinitely often in $e$ with the probability 1. Hence, the legitimate configuration $\gamma_k$ eventually occurs in $e$ with the probability 1, a contradiction. 

{\bf Only If.} Assume that $\PROTO$ is probabilistically self-stabilizing under a distributed randomized scheduler. First, $\PROTO$ satisfies the \emph{strong closure} property. Then, starting from any configuration, there exists at least one execution that converges to a legitimate configuration: $\PROTO$ satisfies the \emph{possible convergence} property. Hence, $\PROTO$ is \emph{weak-stabilizing} and, by Theorem \ref{theo:goudasf}, $\PROTO$ is deterministically self-stabilizing under the Gouda's fairness assumption.
\end{proof}

Theorem \ref{gouda0-2} claims that if the distributed scheduler does not behave as an adversary, then any deterministic weak-stabilizing system stabilizes with a probability 1. So, we could expect that under a synchronous scheduler, which corresponds to a "friendly" behavior of the distributed scheduler, any weak-stabilizing system also stabilizes. Unfortunately, this is not the case: for example, Figure \ref{ex1} (page \pageref{ex1}) depicts a possible synchronous execution of Algorithm \ref{algo:1} that never converges. In contrast, it is easy to see that under a central randomized scheduler, Algorithms \ref{algo:2} and \ref{algo:1} are still probabilistically self-stabilizing (to prove the weak-stabilization of Algorithms \ref{algo:2} and \ref{algo:1} under a distributed scheduler we never use the fact that more that one process can be activated at each step). 
Hence, this means that in some cases, the asynchrony of the system helps its stabilization while the synchrony can be pathological. This could seem unintuitive at first, but this is simply due to the fact that a synchronous scheduler maintains symmetry in the system. However, it is desirable to have a solution that works with both a distributed randomized scheduler and a synchronous one. This is the focus of the following paragraph.

\smallskip

\noindent\textbf{Breaking Synchrony-induced Symetry.} We now propose a simple transformer that permits to break the symetries when the system is synchronous while keeping the convergence property of the algorithm under a distributed randomized scheduler. 
Our transformation method consists in simulating a randomized distributed scheduler when the system behaves in a synchronous way (this method was used in the conflict manager provided in \cite{GT07}): each time an enabled process is activated by the scheduler, it first tosses a coin and then performs the expected action only if the toss returns true. 

In our scheme, we add a new boolean random variable $B_i$ in the code of each processor $i$. We then transform any action $\A :: Guard_{\A} \to \mathtt S_\A$ of the input (deterministic weak-stabilizing) algorithm into the following action $\Trans(\A)$:
\begin{center}
\scriptsize
$\Trans(\A) :: Guard_{\A} \to B_i \gets \Rand_i(true$,$false)$; {\bf if} $B_i$ {\bf then} $\mathtt S_\A$
\end{center} 

Of course, our method does not absolutely forbid synchronous behavior of the system: at any step, there is a strictly positive probability that every enabled process is activated and wins the toss. Such a property is very important because some deterministic weak-stabilizing algorithms under a distributed scheduler require some "synchronous" steps to converge. Such an exemple is provided below.

Consider a network consisting of two neighboring processes, $p$ and $q$, having a boolean variable $B$ and executing the following algorithm:

\begin{algorithm}[H]\caption{Code for a process $i$\label{algo:3}}
\scriptsize
{\bf Input:} $j$: the neighbor of $i$\\
{\bf Variable:} $B_i$: boolean\\
{\bf Actions:}\\
\begin{tabular}{llll}
$\A_1$ & $(\neg B_i \wedge \neg B_j)$ & $\to$ & $B_i \gets true$\\
$\A_2$ & $(B_i \wedge \neg B_j)$      & $\to$ & $B_i \gets false$
\end{tabular}
\end{algorithm}

Trivially, Algorithm \ref{algo:3} is deterministically weak-stabilizing under a distributed strongly fair scheduler for the following predicate: $(B_p \wedge B_q)$. Indeed, if $(B_p$,$B_q)$ $=$ $(true$,$false)$ or $(false$,$true)$, then in the next configuration, $(B_p$,$B_q)$ $=$ $(false$,$false)$ and from such a configuration, three cases are possible in the next step: \emph{(i)} only $B_p \gets true$, \emph{(ii)} only $B_q \gets true$, or \emph{(iii)} $(B_p$,$B_q)$ $\gets$ $(true$,$true)$. In the two first cases, the system retreives a configuration where $(B_p$,$B_q)$ $=$ $(true$,$false)$ or $(false$,$true)$. In the latter case, the system reaches a terminal configuration where $(B_p \wedge B_q)$ holds. Hence, Algorithm \ref{algo:3} requires to converge that $p$ and $q$ move simultaneously when $(B_p$,$B_q)$ $=$ $(false$,$false)$. The transformed version of Algorithm \ref{algo:3} trivially converges with the probability 1 under a distributed randomized scheduler as well as a synchronous one because while the system is not in a terminal configuration, the system regulary passes by the configuration $(B_p$,$B_q)$ $=$ $(false$,$false)$ and from such a configuration, there is a strictly positive probability that both $p$ and $q$ executes $B \gets true$ in the next step.

\smallskip

\noindent\textbf{Transformer Correctness.} Below we prove that our method transforms any deterministic weak-stabilizing system for a distributed scheduler with a finite number of configurations into a randomized self-stabilizing system for a synchronous scheduler. The proof that the transformed system remains a probabilistically self-stabilizing under a randomized scheduler is (trivially) similar and is omitted from the presentation.

Let $\Syst_\Det = (\C_\Det$,$\mapsto_\Det)$ be a system that is deterministically weak-stabilizing for the specification $\Spec$ under a distributed scheduler and having a finite number of configurations. Let $\Legi_\Det \subseteq \C_\Det$ be the (non-empty) set of legitimate configurations of $\Syst_\Det$. Let $\Syst_\Prob = (\C_\Prob$,$\mapsto_\Prob)$ be the probabilistic system obtained by transforming $\Syst_\Det$ according to the above presented method. By construction, any variable $v$ of $\Syst_\Det$ also exists in $\Syst_\Prob$. So, let us denote by $\gamma_{|\Syst_\Det}$ the projection of the configuration $\gamma \in \C_\Prob$ on the variables of $\Syst_\Det$. By Definition, $\forall \gamma \in \C_\Prob$, $\gamma_{|\Syst_\Det} \in \C_\Det$ and $\forall \alpha \in \C_\Det$, $\exists \gamma \in \C_\Prob$ such that $\gamma_{|\Syst_\Det} = \alpha$.

\begin{definition}
Let $\Legi_\Prob = \{\gamma \in \C_\Prob : \gamma_{|\Syst_\Det} \in \Legi_\Det\}$.
\end{definition}

\begin{lemma}[Strong Closure]\label{lem:scprob} Any synchronous execution of $\Syst_\Prob$ starting from a configuration of $\Legi_\Prob$ always satisfies $\Spec$.
\end{lemma}
\begin{proof}
By Definition, $(\Legi_\Prob \neq \emptyset)$ and $(\forall \gamma \in \Legi_\Prob$, $\gamma_{|\Syst_\Det} \in \Legi_\Det)$, \emph{i.e.}, the projection of any configuration of $\Legi_\Prob$ on the variables of $\Syst_\Det$ is a legitimate configuration of $\Syst_\Det$. So, it remains to show that any configuration $\gamma \in \Legi_\Prob$ satisfies the predicate $\mathtt P \equiv (\forall \gamma' \in \C_\Prob : \gamma \mapsto_\Prob \gamma'$, $\gamma' \in \Legi_\Prob)$.

Consider any configuration $\gamma \in \Legi_\Prob$.
\begin{itemize}
\item[-] If $\gamma$ is a \emph{terminal configuration} (\emph{i.e.}, there is no configuration $\gamma' \in \C_\Prob$ such that $\gamma \mapsto_\Prob \gamma'$), then $\gamma$ trivially satisfies $\mathtt P$.  
\item[-] Assume now that $(\exists \gamma' \in \C_\Prob : \gamma \mapsto_\Prob \gamma')$. Consider then any transition $\gamma \mapsto_\Prob \gamma'$. In this transition, every enabled process $p$ executes its enabled action $\Trans_p(\A)$ (the execution is synchronous). First, any $p$ tosses a coin ($B_p \gets \Rand_p(true$,$false)$). Then, two cases are possible:
\begin{itemize}
\item[-] \emph{If every process $p$ looses the toss} (\emph{i.e.}, $\Rand_p(true$,$false)$ returns $true$ for any $p$), then no assignment is performed on the variables that are commun to $\Syst_\Prob$ and $\Syst_\Det$. As a consequence, $\gamma'_{|\Syst_\Det} = \gamma_{|\Syst_\Det}$and, trivially, we have $\gamma' \in \Legi_\Prob$.
\item[-] \emph{If some processes win the toss}, then we can remark that any assignment of a variable commun to $\Syst_\Prob$ and $\Syst_\Det$ performed by Action $\Trans_p(\A)$ exists in Action $\A_p$. Now, $\Syst_\Det$ satisfies the strong closure property for the set $\Legi_\Det$ under a distributed scheduler. So, $\gamma'_{|\Syst_\Det} \in \Legi_\Det$, \emph{i.e.}, $\gamma' \in \Legi_\Prob$.
\end{itemize}
Hence, for any transition $\gamma \mapsto_\Prob \gamma'$, we have $(\gamma \in \Legi_\Prob) \Rightarrow (\gamma' \in \Legi_\Prob)$, \emph{i.e.}, $\gamma$ satisfies $\mathtt P$.
\end{itemize}
\end{proof}

As we assume that $\C_\Det$ is finite and the variables of $\Syst_\Prob$ and $\Syst_\Det$ differ by just a boolean, the following observation is obvious: 

\begin{observation}\label{obs:finite} $\C_\Prob$ is a finite set.
\end{observation}

\begin{lemma}\label{lem:reachLC}
$\forall \gamma \in \C_\Prob$, $\exists \gamma' \in \Legi_\Prob$, $\gamma \leadsto \gamma'$ under a synchronous scheduler.
\end{lemma}
\begin{proof}
Let $\gamma_0 \in \C_\Prob$. Consider the configuration $\alpha_0$ such that $\gamma_{0|\Syst_\Det} = \alpha_0$. By Definition, there exists an execution of $\Syst_\Det$: $\alpha_0$, \dots, $\alpha_k$ such that $\alpha_k \in \Legi_\Det$. Now, for any execution $\alpha_0$, \dots, $\alpha_k$ of $\Syst_\Det$ there exists a corresponding execution of $\Syst_\Prob$: $\gamma_0$, \dots, $\gamma_k$ such that $\forall i \in [1\dots k]$,  $\gamma_{i|\Syst_\Det} = \alpha_i$. Indeed:
\begin{itemize}
\item[(1)] The set of enabled processes is the same in $\alpha_{i-1}$ and $\gamma_{i-1}$, and
\item[(2)] Any step $\gamma_{i-1} \mapsto \gamma_i$ is performed if the subset of enabled processes that win the toss during $\gamma_{i-1} \mapsto \gamma_i$ is exactly the subset of enabled processes that are chosen by the distributed scheduler in $\alpha_{i-1} \mapsto \alpha_i$.
\end{itemize}
Since, $\gamma_{k|\Syst_\Det} = \alpha_k$ and $\alpha_k \in \Legi_\Det$, we have $\gamma_k \in \Legi_\Prob$ and the lemma is proven.  
\end{proof}

\begin{lemma}[Probabilistic Convergence]\label{lem:pcprob} Starting from any configuration, any syn\-chro\-nous execution of $\Syst_\Prob$ reaches a configuration of $\Legi_\Prob$ with the probability 1.
\end{lemma}
\begin{proof}
Consider, by the contradiction, that there exists an execution $e$ of $\Syst_\Prob$ that do not reach any a configuration of $\Legi_\Prob$ with the probability 1. Then, by Lemma \ref{lem:reachLC}, while the system is not in a legitimate configuration it is not in a terminal configuration and, as a consequence, $e$ is infinite. Moreover, as the number of possible configurations of the system is finite (Observation \ref{obs:finite}), there is a subset of configurations $\mathtt W \subset \C_\Prob \setminus \Legi_\Prob$ that appears infinitely often in $e$. By Lemma \ref{lem:reachLC} again, there is two configuration $\gamma \in \mathtt W$ and $\gamma' \in \C_\Prob \setminus \mathtt W$ such that $\gamma \mapsto_\Prob \gamma'$ but the step $\gamma \mapsto_\Prob \gamma'$ never appears in $e$. As execution is synchronous, every enabled process executes an action from $\gamma$ and depending on the tosses, there is a strictly positive probability that the step $\gamma \mapsto_\Prob \gamma'$ occurs from $\gamma$. Now, as $\gamma$ appears infinitely often in $e$, the step $\gamma \mapsto_\Prob \gamma'$ is performed after a finite number of occurences of $\gamma$ in $e$ with the probability 1, a contradiction.
\end{proof}

By Lemmas \ref{lem:scprob} and \ref{lem:pcprob}, we get:

\begin{theorem}\label{theo:WDtoSS}
Assuming a synchronous scheduler, $\Syst_\Prob$ is a \emph{probabilistic self-stabilizing} system for $\Spec$.
\end{theorem}

Using the same approach as for Theorem \ref{theo:WDtoSS}, the following result is straighforward.

\begin{theorem}\label{theo:dp}
Assuming a distributed randomized scheduler, $\Syst_\Prob$ is a \emph{probabilistic self-stabilizing} system for $\Spec$.
\end{theorem}

\section{Conclusion}\label{sect:ccl}

Weak-stabilization is a variant of self-stabilization that only requires the \emph{possibility} of convergence, thus enabling to solve problems that are otherwise impossible to solve with self-stabilizing guarantees. As seen throughout the paper, weak-stabilizing protocols are much easier to design and prove than their self-stabilizing counterparts. Yet, the main result of the paper is the practical impact of weak-stabilization: all deterministic weak-stabilizing algorithms can automatically be turned into probabilistic self-stabilizing ones, provided the scheduling is probabilistic (which is indeed the case for practical purposes). Our approach removes the burden of designing and proving probabilistic stabilization by algorithms designers, leaving them with the easier task of designing weak stabilizing algorithms.

Although this paper mainly focused on the theoretical power of weak-stabilization, a goal for future research is the quantitative study of weak-stabilization, evaluating the expected stabilization time of transformed algorithms.  

\bibliographystyle{plain}
\bibliography{bibliothese}

\begin{thebibliography}{10}

\bibitem{A80}
D.~Angluin.
\newblock Local and global properties in networks of processes.
\newblock In {\em 12th Annual ACM Symposium on Theory of Computing}, pages
  82--93, April 1980.

\bibitem{DBLP:conf/podc/BeauquierGK98}
Joffroy Beauquier, Christophe Genolini, and Shay Kutten.
\newblock {\it }-stabilization of reactive tasks.
\newblock In {\em PODC}, page 318, 1998.

\bibitem{DBLP:journals/dc/BeauquierGJ07}
Joffroy Beauquier, Maria Gradinariu, and Colette Johnen.
\newblock Randomized self-stabilizing and space optimal leader election under
  arbitrary scheduler on rings.
\newblock {\em Distributed Computing}, 20(1):75--93, 2007.

\bibitem{DBLP:journals/siamcomp/BruellGKP99}
Steven~C. Bruell, Sukumar Ghosh, Mehmet~Hakan Karaata, and Sriram~V. Pemmaraju.
\newblock Self-stabilizing algorithms for finding centers and medians of trees.
\newblock {\em SIAM J. Comput.}, 29(2):600--614, 1999.

\bibitem{BH90}
F.~Buckley and F~Harary.
\newblock {\em Distance in Graphs}.
\newblock Addison-Wesley Publishing Compagny, Redwood City, CA, 1990.

\bibitem{BGM89}
J.~Burns, M.~Gouda, and R.~Miller.
\newblock On relaxing interleaving assumptions.
\newblock {\em Proceedings of the MCC Workshop on Self-Stabilizing Systems,
  Austin, Texas}, 1989.

\bibitem{BGM93}
James~E. Burns, Mohamed~G. Gouda, and Raymond~E. Miller.
\newblock Stabilization and pseudo-stabilization.
\newblock {\em Distrib. Comput.}, 7(1):35--42, 1993.

\bibitem{DBLP:conf/sss/DasguptaGX07}
Anurag Dasgupta, Sukumar Ghosh, and Xin Xiao.
\newblock Probabilistic fault-containment.
\newblock In {\em Stabilization, Safety, and Security of Distributed Systems,
  9th International Symposium, SSS}, volume 4838 of {\em Lecture Notes in
  Computer Science}, pages 189--203. Springer, 2007.

\bibitem{DGT04j}
Ajoy~K. Datta, Maria Gradinariu, and Sébastien Tixeuil.
\newblock Self-stabilizing mutual exclusion with arbitrary scheduler.
\newblock {\em The Computer Journal}, 47(3):289--298, 2004.

\bibitem{Dij74}
EW~Dijkstra.
\newblock Self stabilizing systems in spite of distributed control.
\newblock {\em Communications of the Association of the Computing Machinery},
  17:643--644, 1974.

\bibitem{citeulike:976108}
Shlomi Dolev.
\newblock {\em Self-Stabilization}.
\newblock {The MIT Press}, March 2000.

\bibitem{GT02c}
Christophe Genolini and Sébastien Tixeuil.
\newblock A lower bound on $k$-stabilization in asynchronous systems.
\newblock In {\em Proceedings of {IEEE} 21st Symposium on Reliable Distributed
  Systems ({SRDS}'2002)}, Osaka, Japan, October 2002.

\bibitem{Gouda01}
Mohamed~G. Gouda.
\newblock The theory of weak stabilization.
\newblock In {\em WSS}, pages 114--123, 2001.

\bibitem{GT07}
Maria Gradinariu and S{\'e}bastien Tixeuil.
\newblock Conflict managers for self-stabilization without fairness assumption.
\newblock In {\em 27th IEEE International Conference on Distributed Computing
  Systems (ICDCS)}, page~46. IEEE Computer Society, 2007.

\bibitem{Her92}
T~Herman.
\newblock {Self-stabilization: ramdomness to reduce space}.
\newblock {\em Information Processing Letters}, 6:95--98, 1992.

\bibitem{DBLP:journals/ipl/Herman90}
Ted Herman.
\newblock Probabilistic self-stabilization.
\newblock {\em Inf. Process. Lett.}, 35(2):63--67, 1990.

\bibitem{DBLP:conf/podc/IsraeliJ90}
Amos Israeli and Marc Jalfon.
\newblock Token management schemes and random walks yield self-stabilizing
  mutual exclusion.
\newblock In {\em PODC}, pages 119--131, 1990.

\bibitem{T07c}
Sébastien Tixeuil.
\newblock {\em Wireless Ad Hoc and Sensor Networks}, chapter Fault-tolerant
  distributed algorithms for scalable systems.
\newblock ISTE, October 2007.
\newblock ISBN: 978 1 905209 86.

\end{thebibliography}

\appendix

\section{Proof of Theorem \ref{theo:algo2w}}\label{app:A}

\begin{definition}[$\tokenh$]\label{def:th} Let $\gamma$ be a configuration. Let $\tokenh(\gamma)$ be the set of processes $p$ satisfying $Token(p)$ in the configuration $\gamma$. 
\end{definition}

\begin{definition}[$\LCS$]\label{def:lcs} Let $\LCS$ be the set of configurations $\gamma$ such that $\gamma$ satisfies $|\tokenh(\gamma)|$ $=$ $1$.
\end{definition}

\begin{definition}[$\predpath$]\label{def:predpath}
Let $p$ and $q$ be two distinct processes. We call $\predpath(p,q)$ be the unique path $p_0$, \dots, $p_k$ such that: (1) $p_0 = p$, (2) $\forall i \in [1\dots k]$, $Pred_{p_k} = p_{k-1}$, and $p_k = q$.
\end{definition}

\begin{remark}
Let $p$ and $q$ be two distinct processes. $\predpath(p,q) \neq \predpath(p,q)$.
\end{remark} 

\begin{definition}[$\mtd$: MinTokenDistance]\label{def:mtd} Let $\gamma$ be a configuration such that $\gamma$ satisfies $|\tokenh(\gamma)|$ $>$ 1. We denote by $\mtd(\gamma)$ the length of the shortest path $\predpath(p,q)$ such that $Token(p)$ and $Token(q)$ in $\gamma$.
\end{definition}

\begin{lemma}\label{lem:geq1}
For any configuration $\gamma$, we have $|\tokenh(\gamma)|>0$.
\end{lemma}
\begin{proof}
Assume, by the contradiction, that there is a configuration $\gamma$ such that $|\tokenh(\gamma)|$ $=$ 0. Let $p_0$, \dots, $p_{N-1}$ be an hamiltonian path of processes such that, $\forall i \in [0\dots N-1]$, $p_i = Pred_{p_{(i+1) \bmod N}}$. Then, ($\forall i \in [0\dots N-1]$, $\neg Token(p_i)$) implies that ($\forall i \in [0\dots N-1]$, $[dt_{p_{(i+1) \bmod N}}$ $=$ $((dt_{p_i} + 1)$ $\bmod$ $m_{N})]$) which is not possible because $(N \bmod m_N) \neq 0$, a contradiction.
\end{proof}

\begin{lemma}[Possible Convergence]\label{lem:pc:algo2} Starting from any configuration, there exists at least one possible execution that reaches a configuration $\gamma \in \LCS$.
\end{lemma}
\begin{proof}
Any configuration satisfies $|\tokenh| > 0$ by Lemma \ref{lem:geq1}. Consider any configuration $\gamma$ satisfying $|\tokenh(\gamma)| > 1$. Let us study the two following cases:

\emph{$\mtd(\gamma) = 1$.} In this case, there exists two processes $p$ and $q$ such that $|\predpath(p$,$q)| = 1$, \emph{i.e.}, $p$ is the predecessor of $q$ and both $p$ and $q$ satisfies the predicate $Token$ (\emph{i.e.}, both $p$ and $q$ hold a token). If only $p$ executes Action $\mathtt{A}$ in the next step, then $p$ satisfies $\neg Token$ in the next configuration $\gamma'$ and, as the consequence, $|\tokenh(\gamma')| < |\tokenh(\gamma)|$. 

\emph{$\mtd(\gamma) > 1$.} Let consider two processes $p$ and $q$ such that $|\predpath(p$,$q)| = \mtd(\gamma)$. Then, Action $\mathtt{A}$ is enabled at $p$ and if only $p$ moves in the next step, then $|\predpath(p$,$q)|$ decreases of one unit in the next configuration. Hence, inductively there exists an execution from $\gamma$ that reaches a configuration $\gamma'$ such that $\mtd(\gamma') = 1$. 

Hence, from any configuration $\gamma$ such that $|\tokenh(\gamma)| > 1$ there always exists an execution where the cardinal of $\tokenh$ eventually decreases and the lemma is proven.
\end{proof}

\begin{lemma}[Strong Closure]\label{lem:sc:algo2}
Any execution starting from a configuration $\gamma$ such that $\gamma \in \LCS$ always satisfies the specification of the token circulation. 
\end{lemma}
\begin{proof}To prove this lemma we show that $\forall \gamma \in \LCS$, $\forall \gamma \mapsto \gamma'$, (1) $\gamma' \in \LCS$ ($\LCS$ is closed) and (2) the token holder in $\gamma'$ is the successor of the token holder in $\gamma$. 

Consider a configuration $\gamma$ such that $|\tokenh(\gamma)| = 1$. Let $q$ be the only process satisfying $Token(q)$ in $\gamma$. Let $p$ and $s$ be the predecessor and the successor of $q$ in $\gamma$, respectively. Then, in $\gamma$, $q$ is the only enabled process, $dt_q \neq ((dt_p + 1) \bmod m_{N})$, and $dt_s = ((dt_q + 1) \bmod m_{N})$. During the next step, $q$ executes $\mathtt{A}$ and, as a consequence, $dt_q = ((dt_p + 1) \bmod m_{N})$ and $dt_s \neq ((dt_q + 1) \bmod m_{N})$ in the next configuration $\gamma'$: $s$ is the only token holder in $\gamma'$, which proves the lemma.
\end{proof}

\bigskip

\noindent {\bf Proof of Theorem \ref{theo:algo2w}.} By Lemmas \ref{lem:pc:algo2} and \ref{lem:sc:algo2}, the theorem is obvious. $\Box$

\section{Proof of Theorem \ref{theo:algo1w}}\label{app:B}

\begin{definition}[$\ppath$]\label{def:ppath} We call $\ppath(p)$ the \emph{unique maximal} path $p_0$, \dots, $p_k$ such that: (1) $p_k = p$, (2) $\forall i \in [1\dots k]$, $Par_i = p_{i-1}$, and (3) $p_0$ satisfies $[(Par_{p_0} \neq \perp) \Rightarrow (Par_{Par_{p_0}} = p_0)]$.
\end{definition}

\begin{notation}
Let $p$ be a process. In the following, we denote by $\ROOT(p)$ the \emph{initial extremity} of $\ppath(p)$ (\emph{n.b.}, $(Par_p = \perp) \Rightarrow (\ROOT(p) = p)$).
\end{notation}

\begin{remark}
As the network is acyclic, for any process $p$, $\ppath(p)$ has a finite length.
\end{remark}

\begin{definition}[$\LC$]\label{def:lc}
Any configuration $\gamma$ satisfies the predicate $\LC(\gamma)$ \emph{iff} the two following conditions hold in $\gamma$: (1) there exists exactly one process $p$ that satisfies $Par_p = \perp$ and (2) for any process $q \neq p$, $\ROOT(q) = p$.
\end{definition}

\begin{remark}\label{rem:one}
There is exactly one process satisfying $\leader$ in any configuration $\gamma$ satisfying $\LC(\gamma)$.
\end{remark}

\begin{lemma}\label{lem:1Lreach}
In any configuration where every process satisfies $\neg \leader$, there exists at least one process $p$ such that Action $\mathtt{A_1}$ is enabled at $p$. 
\end{lemma}
\begin{proof}
Let $NearestCenter(p)$ be the center process at the smallest distance from the process $p$. Let $\mathcal{DNC}_{max} = \lceil D/2 \rceil$ be the maximal distance between any process $p$ and $NearestCenter(p)$. Let $\mathcal{DNC}^{-1}(p) = \mathcal{DNC}_{max}-d(p,NearestCenter(p))$.

 Assume, by the contradiction, that there exists a configuration $\gamma$ where every process satisfies $\neg \leader$ and no Action $\mathtt{A_1}$ is enabled. We show the contradiction in two steps:

{\bf Step 1.} First, we prove that any process $p$ such that $\mathcal{DNC}^{-1}(p) = d$ with $0 \leq d <  \mathcal{DNC}_{max}$ (actually the non-center processes) satisfies $Par_p = q$ in $\gamma$ with $\mathcal{DNC}^{-1}(q) = d + 1$.

{\bf Step 2.} Then, we show the contradiction using {\bf Step 1}.

\smallskip

\noindent {\bf Step 1.} (by induction)

\emph{Induction for d = 0.} By Definition, any process $p$ such that $\mathcal{DNC}^{-1}(p) = 0$ is a leaf node. As $p$ satisfies $\neg \leader(p)$, $Par_p = q$ holds in $\gamma$ where $q$ is the only neighbor of $p$. Now, by definition, $\mathcal{DNC}^{-1}(q) = \mathcal{DNC}^{-1}(p) + 1 = 1$. Hence, the induction holds for $d = 0$.

\emph{Induction Assumption:} Let $k \in [0\dots \mathcal{DNC}_{max}-1]$. Assume that any process $p$ such that $0 \leq \mathcal{DNC}^{-1}(p) < k$ satisfies $Par_p = q$ in $\gamma$ with $\mathcal{DNC}^{-1}(q) = k + 1$.

\emph{Induction for d = k + 1.} Consider a process $p$ such that $\mathcal{DNC}^{-1}(p) = k + 1$. Then, $\mathcal{DNC}^{-1}(p) < \mathcal{DNC}_{max}$ and, by definition, $p$ has one neighbor $q$ such that $\mathcal{DNC}^{-1}(q) = k+2$ and all its other neighbors $q'$ satisfies $\mathcal{DNC}^{-1}(q') = k$. Assume, by the contradiction, that $Par_q = v$ with $\mathcal{DNC}^{-1}(v) = k$. Then, any other $v$'s neighbor, $v'$, satisfies $\mathcal{DNC}^{-1}(v) = k-1$. Hence, by induction assumption, any process $v'$ satisfies $Par_{v'} = v$. Now, $Par_v \neq \perp$ because $v$ satisfies $\neg \leader(v)$. So, Action $\mathtt{A_1}$ is enabled at $v$, a contradiction. Hence, $Par_p = q$ where $q$ is the only neighbor of $p$ such that $\mathcal{DNC}^{-1}(q) = k+2$ and the induction holds for $d = k+1$.

\smallskip

\noindent {\bf Step 2.}

We now show the contradiction. By Property \ref{prop:center} (page \pageref{prop:center}), we can split our study in the two following cases:

\emph{There is one center $c$ in the network.} In this case, any neighbor of $c$, $c'$, satisfies $\mathcal{DNC}^{-1}(c') = \mathcal{DNC}_{max} - 1$. In this case, any process $c'$ also satisfies $Par_{c'} = c$ ({\bf Step 1}). Now, $Par_c \neq \perp$ because $c$ satisfies $\neg \leader(c)$. So, Action $\mathtt{A_1}$ is enabled at $c$, a contradiction. 

\emph{There is two neighboring centers $c_0$ and $c_1$ in the network.} In this case, any non-center neighbor of $c_i$ ($i \in \{0,1\}$), $c'_i$, satisfies $\mathcal{DNC}^{-1}(c'_i) = \mathcal{DNC}_{max} - 1$. In this case, any process $c_i'$ also satisfies $Par_{c_i'} = c_i$ ({\bf Step 1}). Assume now, by the contradiction, that one the centers $c_i$ ($i \in \{0,1\}$) satisfies $Par_{c_i} = c'_i$ where $c'_i$ is a neighbor such that $\mathcal{DNC}^{-1}(c'_i) = \mathcal{DNC}_{max} - 1$. Then, any other $c'_i$'s neighbor also satisfies $Par = c'_i$ ({\bf Step 1}). Now, $Par_{c'_i} \neq \perp$ because $c'_i$ satisfies $\neg \leader(c'_i)$. So, Action $\mathtt{A_1}$ is enabled at $c'_i$, a contradiction. Hence, $Par_{c_0} = c_1$ and $Par_{c_1} = c_0$ and Action $\mathtt{A_1}$ is both enabled at $c_0$ and $c_1$, a contradiction.
\end{proof}

\noindent The following corollary simply holds by the fact that after executing Action $\mathtt{A_1}$, a process satisfies $\leader$.

\begin{corollary}\label{coro:1Lreach}
Starting from any configuration, the system can reach in at most one step a configuration where at least one process satisfies $\leader$.
\end{corollary}

\begin{lemma}\label{lem:converge}
From any configuration where at least one process satisfies $\leader$, there is a possible execution that reaches a configuration $\gamma$ satisfying $\LC(\gamma)$.
\end{lemma}
\begin{proof}Let $p$ be a process satisfying $\leader(p)$. Let $Tree(p) = \{q \in V$, $\ROOT(q) = p\}$. First, from Definition \ref{def:lc}, we can trivially deduce that a configuration satisfies $\LC$ \emph{iff} it contains a unique tree $Tree(p)$ such that $Tree(p) = V$.

Consider then a configuration $\gamma$ satisfying $\neg \LC(\gamma)$ where there exists a process $p$ satisfying $\leader(p)$. So, $Tree(p) \subset V$. Let $NonTree(p) = V \setminus Tree(p)$. To prove this lemma, we just show below that from such a configuration $\gamma$ is always possible to reach (in a finite number of step) a configuration $\gamma'$ where the cardinal of $NonTree(p)$ decreased.

First, $p$ satisfying $\leader(p)$ in $\gamma$, so, $Tree(p) \neq \emptyset$ in $\gamma$. Then, as $\gamma$ satisfies $\neg \LC(\gamma)$, $NonTree(p) \neq \emptyset$ and, as the network is connected, there two neighboring processes $v$ and $w$ such that $v \in Tree(p)$ and $w \in NonTree(p)$ in $\gamma$. Also, $Par_v \neq w$ and $Par_w \neq v$ in $\gamma$ by Definition \ref{def:ppath}. Consider then the two following cases:
\begin{itemize} 
\item[-] \emph{$Par_w \neq \perp$ in $\gamma$.} In this case, Action $\mathtt{A_2}$ is enabled at $w$ until (at least) $Par_w = v$. Now, after at most $\Delta_w - 1$ executions of Action $\mathtt{A_2}$, $Par_w$ points out to $v$. Hence, if only actions $\mathtt{A_2}$ at $w$ are executed until $Par_w$ points out to $v$, there is an execution from $\gamma$ that reaches a configuration $\gamma'$ where $|NonTree(p)|$ decreases of one unit. 
\item[-] \emph{$Par_w = \perp$ in $\gamma$.} In this case, as $Par_v \neq w$, Action $\mathtt{A_3}$ is enabled at $w$. If only $w$ moves in the next step, then either (1) $Par_w$ points out to $v$ in the next configuration and $|NonTree(p)|$ decreases of one unit, or (2) $Par_w \notin \{v$,$\perp\}$ in the next configuration and we retreive the previous case.
\end{itemize}
Hence, from any configuration $\gamma$ satisfying $\neg \LC(\gamma)$ where there is a process $p$ satisfying $\leader(p)$, it is always possible to reach a configuration $\gamma'$ where $|NonTree(p)|$ decreased.
\end{proof}

\noindent By Corollary \ref{coro:1Lreach} and Lemma \ref{lem:converge}, follows:

\begin{lemma}[Possible Convergence]\label{lem:converge2}
Starting from any configuration, there exists at least one possible execution that reaches a configuration $\gamma$ satisfying $\LC(\gamma)$.
\end{lemma}

\begin{lemma}[Strong Closure]\label{lem:tc}
Let $\gamma$ be a configuration. $\gamma$ satisfies $\LC(\gamma)$ \emph{iff} $\gamma$ is a \emph{terminal configuration}.
\end{lemma}
\begin{proof}~

{\bf If.} Consider a configuration $\gamma$ satisfying $\LC(\gamma)$. Let $p$ be the only process that satisfies $Par = \perp$ in $\gamma$. By Definition \ref{def:lc}, any neighbor $p'$ of $p$ satisfies $Par_{p'} = p$ and, as a consequence, $p$ is disabled. Consider now any process $q$ such that $Par_q \neq \perp$. As we are in a tree network, there is only one path linking any process $q$ to $p$, so, by Definition \ref{def:lc}, any process $q$ points out with $Par_q$ the unique neighbor $q'$ whereby it can reach $p$, $q'$ does not point out to $q$ with $Par_{q'}$, and all other neighbors of $q$ points out to $q$ with their $Par$ pointer. As a consequence, any process $q$ is disabled. Hence, $\gamma$ is a terminal configuration.

{\bf Only If.} (by the contraposition) By Lemma \ref{lem:converge2}, any configuration $\gamma$ satisfying $\neg \LC(\gamma)$ is not \emph{terminal}.
\end{proof}

\bigskip

\noindent {\bf Proof of Theorem \ref{theo:algo1w}.} Follows from lemmas \ref{lem:converge2} and \ref{lem:tc}, and Remark \ref{rem:one}. $\Box$

\end{document}